\documentclass[aps,prd,reprint,superscriptaddress,nofootinbib,longbibliography]{revtex4-2}

\usepackage{amsmath,amssymb,bm,mathtools}
\usepackage{microtype}
\usepackage[colorlinks=true,citecolor=blue,linkcolor=blue,urlcolor=black]{hyperref}

\newcommand{\dd}{\mathrm d}
\newcommand{\ii}{\mathrm i}
\newcommand{\Tr}{\operatorname{Tr}}
\newcommand{\lp}{\ell_{\rm P}}
\newcommand{\tG}{\widetilde{\mathbf G}}
\newcommand{\tg}{\widetilde g}
\newcommand{\tR}{\widetilde{\bm{\mathcal R}}}
\newcommand{\tM}{\widetilde{\mathbf M}}
\newcommand{\tTheta}{\widetilde{\bm\Theta}}
\newcommand{\tI}{\widetilde{\mathbf I}}
\newcommand{\Wresp}{\widetilde{\bm{\mathcal W}}}
\newcommand{\Gfield}{\widetilde{\bm{\mathcal G}}}
\newcommand{\Fcal}{\mathcal F}
\newcommand{\Dcal}{\mathcal D}
\newcommand{\Order}{\mathcal O}
\newcommand{\ket}[1]{\lvert #1\rangle}
\newcommand{\bra}[1]{\langle #1\rvert}
\newcommand{\RicTF}{\mathring R}

\begin{document}

\title{Gravitational Waves and Matter Constraints in Gravity from Entropy}

\author{David S. Pereira}
\email{djpereira@ciencias.ulisboa.pt}
\affiliation{Departamento de F\'{i}sica, Faculdade de Ci\^{e}ncias da Universidade de Lisboa, Campo Grande, Edif\'{i}cio C8, P-1749-016 Lisbon, Portugal}
\affiliation{Instituto de Astrof\'{i}sica e Ci\^{e}ncias do Espa\c{c}o, Faculdade de Ci\^{e}ncias da Universidade de Lisboa, Campo Grande, Edif\'{i}cio C8, P-1749-016 Lisbon, Portugal}

\begin{abstract}
We show that Gravity from Entropy (GfE) exhibits a tensor-propagation pathology on generic FLRW backgrounds. In vacuum, in the symmetry-inheriting single-multiplet sector, and for the algebraic perfect-fluid constitutive embedding, the strict-principal TT system has a nonzero real characteristic only when \(\dot H=K/a^2\). Away from this locus the temporal principal roots become complex, signaling a high-frequency hyperbolicity obstruction. Under the stated isotropy assumptions, a rank theorem further shows that one foundational multiplet cannot provide the isotropic bivector response needed to remove this problem.
\end{abstract}

\maketitle

\emph{Introduction.---}
 A viable candidate theory of modified or quantum gravity must reproduce the successful phenomenology of general relativity while also possessing a consistent perturbative and causal structure. This distinction is particularly relevant for approaches in which gravitational dynamics emerge from thermodynamic or information-theoretic principles. The connection between gravity and entropy originates in black-hole thermodynamics~\cite{Bekenstein:1973ur,Hawking:1975vcx} and has since been developed into thermodynamic and entanglement-based routes to Einstein equations~\cite{Jacobson:1995ab,Ryu:2006bv,VanRaamsdonk:2010pw,Faulkner:2013ica,Jafferis:2015del,Jacobson:2015hqa,Dorau:2025hmq}. More generally, effective and modified theories of gravity provide many examples in which agreement with GR at the background or infrared level does not guarantee stable, hyperbolic, or causal propagation of perturbations~\cite{Donoghue:1994dn,Reall:2014pwa,Papallo:2017qvl,Escamilla-Rivera:2012jfb,Thaalba:2026eyv}.

Gravity from Entropy (GfE) makes this distinction concrete~\cite{Bianconi:2024hts,Bianconi:2024aju}. Its trace-log action acts on a scalar--one-form--two-form multiplet, with matter and curvature entering a common form-space operator while Ricci and Riemann curvature occupy different blocks~\cite{Bianconi:2024aju}. Gravitational waves can therefore probe response information hidden from homogeneous FLRW equations. We derive the tensor characteristic directly from the foundational trace-log theory and subsequently prove exact agreement with its constrained auxiliary G-field formulation. This question is timely because GfE has already been applied to inflation, black holes, thermodynamics, area-law and diffusion phenomena, and the Minkowski spectrum~\cite{Bianconi:2025rnd,Bianconi:2025eok,Thattarampilly:2025krv,Bianconi:2025awa,Thattarampilly:2026wsw,Pereira:2026mjk}.

We find that strict-principal tensor causality imposes a stronger condition than the FLRW background equations. In the sectors studied here, a real strict-principal tensor characteristic exists only when the distinct curvature-response channels reconstruct the same Lorentzian cone, which then coincides with the metric light cone. Two independent matter tests give the same obstruction. The direct isotropic perfect-fluid operator used in GfE thermodynamics~\cite{Bianconi:2025awa}, treated as the algebraic constitutive embedding used in low-energy GfE cosmology, cannot compensate the relevant curvature splitting despite reproducing Friedmann background dynamics in that regime. For one foundational multiplet, full Douglis--Nirenberg principal isotropy ensures tensor-sector closure; separately, a rank theorem shows that independent rotational invariance of the composite spatial one-form and bivector responses, together with the rank-two bound, forces the relevant responses to vanish. Vacuum, the low-energy perfect-fluid embedding, and one-multiplet backgrounds satisfying these conditions therefore select pointwise-Einstein FLRW backgrounds at strict-principal order. The flat vacuum slow-roll and phantom-like solutions~\cite{Thattarampilly:2025krv} consequently fail this strict-principal tensor condition away from exact de Sitter evolution.

The obstruction is stronger than a modified gravitational-wave speed: away from response matching the degree-four TT symbol has no nonzero real characteristic covector. Causality constraints on ultraviolet particle content are well known~\cite{Adams:2006sv,Camanho:2014apa}, as are nonmetric characteristics in modified gravity~\cite{Schuller:2009hn,Reall:2021voz,Hollands:2026aeh,Dittrich:2026acv}. The mechanism here is different: inequivalent form-space curvature responses must reconstruct a common cone, turning tensor propagation into a finite-rank representation constraint on microscopic matter.

\emph{Foundational GfE formulation and FLRW responses.---}
We use $\hbar=c=1$, signature $(-,+,+,+)$, and the corrected conventions of the foundational continuum theory~\cite{Bianconi:2024aju}. A bar denotes a background quantity and $\delta$ a linear perturbation; $\lp$ is the Planck length. The dynamical matter and geometry belong to
\begin{equation}
 \Fcal=\Lambda^0\oplus\Lambda^1\oplus\Lambda^2.
  \label{eq:Fspace}
\end{equation}
The matter state is given by
\begin{equation}
 \ket{\Phi}=\phi\oplus\omega_\mu\dd x^\mu
 \oplus\zeta_{\mu\nu}\dd x^\mu\wedge\dd x^\nu ,
\end{equation}
where $\phi$, $\omega_\mu$, and $\zeta_{\mu\nu}$ are the scalar, one-form, and two-form components. The spacetime metric induces the block-diagonal reference metric
$\tg=1\oplus g_{(1)}\oplus g_{(2)}$ on $\Fcal$, with $g_{(1)}=g_{\mu\nu}$ and $g_{(2)}$ the induced bivector metric. Matter and curvature deform it to the second form-space metric~\cite{Bianconi:2024aju}
\begin{equation}
 \tG=\tg+\alpha\tM-\beta\tR,
 \qquad
 \alpha=\alpha'\lp^4,\qquad \beta=\beta'\lp^2,
 \label{eq:Gtop}
\end{equation}
where $\alpha',\beta'>0$ are dimensionless and $\tR=R\oplus R_{(1)}\oplus R_{(2)}$ denotes the scalar, Ricci, and Riemann curvature operators on the three form degrees. Matter enters through
\begin{equation}
 \tM=\Dcal\ket{\Phi}\bra{\Phi}\Dcal
 +(m^2+\xi R)\ket{\Phi}\bra{\Phi},
 \quad
 \Dcal=P_{\Fcal}(d+\delta_D)P_{\Fcal},
 \label{eq:Mdef}
\end{equation}
with bosonic mass parameter $m$ and nonminimal coupling $\xi$; $P_{\Fcal}$ projects onto the truncated form space, while $d$ and $\delta_D$ are the exterior derivative and coderivative. The local outer products are degree preserving, so
$\tM=M_{(0)}\oplus M_{(1)}\oplus M_{(2)}$.

Raising the second form-space index with $\tg^{-1}$ defines the composite mixed operator
\begin{equation}
 \tTheta\equiv\tG\tg^{-1}
 =\tI+\alpha\tM\tg^{-1}-\beta\tR\tg^{-1},
 \label{eq:ThetaDef}
\end{equation}
that is the core quantity in the foundational action of GfE~\cite{Bianconi:2024aju}
\begin{equation}
 S_{\rm ent}[g,\Phi]
 =-\lp^{-4}\!\int\!\dd^4x\sqrt{-g}\,
 \Tr_{\Fcal}\ln\tTheta ,
 \label{eq:originalAction}
\end{equation}
where $\Tr_{\Fcal}$ is the trace over $\Lambda^0\oplus\Lambda^1\oplus\Lambda^2$. At this stage the independent fields are only $g_{\mu\nu}$ and $\Phi$. We work on the finite positive-spectrum branch, for which the background operator $\bar\tTheta$ is real, positive, and invertible, and define the \emph{algebraic inverse response}
\begin{equation}
 \Wresp\equiv\tTheta^{-1}.
 \label{eq:Wresponse}
\end{equation}
Thus $\Wresp$ is a composite of the foundational fields, not an additional degree of freedom. Every physical result obtained is derived from Eqs.~\eqref{eq:Mdef}--\eqref{eq:originalAction}; the auxiliary fields matching will be done in the end.

We now specialize to an FLRW background:
\begin{equation}
 \dd s^2=-\dd t^2+a^2(t)\gamma_{ij}^{(K)}\dd x^i\dd x^j,
 \qquad H\equiv\dot a/a ,
 \label{eq:FLRW}
\end{equation}
with spatial-curvature parameter $K$. At an event, spatial rotations decompose the spatial one-form sector as a triplet $\mathbf3_s$ and the bivector space as
$\Lambda^2\simeq\mathbf3_E\oplus\mathbf3_B$, where $E$ labels time--space planes and $B$ purely spatial planes. Rotational invariance fixes the spatial one-form response to be proportional to $I_3$, while $SO(3)$ isotropy alone permits mixing between the two equivalent bivector triplets, with the general form $A_{\mathcal W}\otimes I_3$, where $A_{\mathcal W}$ is a $2\times2$ matrix. In the vacuum, symmetry-inheriting single-multiplet, and perfect-fluid sectors considered below, however, $M_{(2)}=0$ and the FLRW curvature operator is $E/B$ diagonal, so we may write
\begin{equation}
 \bar\Wresp_{(1)}\big|_{\mathbf3_s}=\mathcal W_s I_3,
 \quad
 \bar\Wresp_{(2)}
 =\operatorname{diag}(\mathcal W_E I_3,\mathcal W_B I_3),
 \label{eq:responseEigenvalues}
\end{equation}
which defines the three TT-relevant inverse-response eigenvalues
$\mathcal W_s,\mathcal W_E,\mathcal W_B$; here $I_3$ is the identity on a rotational triplet.

For vacuum, all matter blocks vanish identically, and for the symmetry-inheriting foundational state
$\bar\Phi=\phi(t)\oplus\omega_0(t)\dd t\oplus0$,
\begin{equation}
 \Dcal\ket{\bar\Phi}
 =(\dot\omega_0+3H\omega_0)\oplus\dot\phi\,\dd t\oplus0 .
 \label{eq:symmetryD}
\end{equation}
Hence $\bar M_{(2)}=0$ and the spatial one-form matter block vanishes, even for complex homogeneous amplitudes. The TT-relevant eigenvalues of $\bar\tTheta$ are then
\begin{align}
 \theta_s&=1-\beta\left(\dot H+3H^2+2K/a^2\right),\nonumber\\
 \theta_E&=1-2\beta(H^2+\dot H),\
 \theta_B=1-2\beta(H^2+K/a^2),
 \label{eq:geometricTheta}
\end{align}
with $\mathcal W_A=\theta_A^{-1}$ for $A=s,E,B$.

The same $E/B$ responses occur in the macroscopic perfect-fluid realization used in low-energy GfE thermodynamics~\cite{Bianconi:2025awa},
\begin{align}
 M^{\rm pf}_{(1)\mu\nu}
 &=\frac18\left[4(\rho+P)U_\mu U_\nu
 +(\rho-P)g_{\mu\nu}\right],\nonumber\\
 M^{\rm pf}_{(0)}&=M^{\rm pf}_{(2)}=0,
 \label{eq:pf}
\end{align}
where $\rho$ and $P$ are the fluid energy density and pressure. We use Eq.~\eqref{eq:pf} only as this algebraic constitutive prescription, not as a microscopic fluid action; uppercase $P$ is distinguished from the wave number $p$ used below. Standard ideal-fluid perturbations carry helicities $0,\pm1$ but no helicity-$\pm2$ matter amplitude, and the operator~\eqref{eq:pf} is algebraic in the metric and fluid variables. In the comoving frame its spatial mixed one-form block is $(\rho-P)I_3/8$, so
\begin{align}
 \theta_s^{\rm pf}
 &=1+\frac{\alpha}{8}(\rho-P)
 -\beta\left(\dot H+3H^2+2K/a^2\right),\nonumber\\
 \theta_E^{\rm pf}&=\theta_E,\quad
 \theta_B^{\rm pf}=\theta_B .
 \label{eq:pfTheta}
\end{align}
Thus the perfect fluid dresses $\mathcal W_s$ but leaves the two bivector responses purely geometric.

For all three cases just described---vacuum, the symmetry-inheriting foundational multiplet, and the perfect-fluid constitutive embedding---define
\begin{equation}
 \Delta_H\equiv\dot H-\frac{K}{a^2}.
 \label{eq:DeltaHdef}
\end{equation}
Since $\theta_B-\theta_E=2\beta\Delta_H$, their common bivector responses satisfy
\begin{equation}
 \mathcal W_E-\mathcal W_B
 =2\beta\,\mathcal W_E\mathcal W_B\,\Delta_H .
 \label{eq:DeltaH}
\end{equation}
This identity is the geometric input to the tensor-cone theorem below.

\emph{Tensor cone matching: vacuum, foundational, and perfect-fluid sectors.---}
We now vary the original trace-log action~\eqref{eq:originalAction} directly. Freeze the FLRW coefficients at an event and take a TT perturbation
$h_{ij}=e^A_{ij}h_Ae^{-\ii\omega t+\ii pz}$, with physical wavenumber $p$, local frequency $\omega$, polarization label $A=+,\times$, and
$e^A_{ij}e^{B\,ij}=2\delta^{AB}$~\cite{Isaacson:1968hbi,Fier:2021fbt}. The associated frozen covector may be taken as
$k_\mu=(-\omega,0,0,p)$. By the \emph{strict-principal} sector we mean the highest-derivative part of the frozen linearized equations~\cite{Sarbach:2012pr,Reall:2021voz}.

Let $\mathsf Y\equiv\delta\tTheta$. The exact second variation of the trace-log factor is
\begin{equation}
 \delta^2[-\Tr\ln\tTheta]
 =\Tr(\bar\Wresp\mathsf Y\bar\Wresp\mathsf Y)
 -\Tr(\bar\Wresp\,\delta^2\tTheta).
 \label{eq:logvariation}
\end{equation}
The second term contains at most two metric derivatives in the quadratic action, whereas the first contains the square of the linearized curvature. In each of the three sectors above, matter contributes no independent two-derivative TT metric insertion: this is immediate in vacuum and for the algebraic perfect-fluid prescription, while for the symmetry-inheriting foundational multiplet it follows from Eq.~\eqref{eq:symmetryD} and the derivative counting summarized in the Appendix. Hence
\begin{align}
 \left.\delta^2\mathcal L_{\rm ent}\right|_{4\partial}
 &=\Tr_{\Fcal}(\bar\Wresp\mathsf Y_{\rm TT}^{(2)}
 \bar\Wresp\mathsf Y_{\rm TT}^{(2)}),\nonumber\\
 \mathsf Y_{\rm TT}^{(2)}
 &=-\beta\,\delta(\tR\tg^{-1})_{\rm TT}^{(2)} .
 \label{eq:Hessian}
\end{align}
The one-form and bivector contractions give the common TT principal eigenvalue
\begin{equation}
 q_4(\omega,p)
 =\frac{\mathcal W_s^2}{2}(\omega^2-p^2)^2
 +2(\mathcal W_E\omega^2-\mathcal W_Bp^2)^2 ,
 \label{eq:q4}
\end{equation}
up to the common factor $\beta^2$. The bivector contribution in Eq.~\eqref{eq:q4} uses the foundational flattened independent-pair trace on $\Lambda^2$, including the $2!$ factor associated with the two-form convention; this fixes the relative coefficient of the Riemann-response term (Appendix). Equation~\eqref{eq:q4} therefore applies to vacuum, to the symmetry-inheriting foundational single-multiplet class, and to the macroscopic perfect-fluid constitutive sector, with the appropriate $\mathcal W_s$ in each case. Rotational invariance together with the symmetry of the quadratic Hessian makes the TT principal matrix proportional to $q_4\delta_{AB}$, so both polarizations share the same characteristic and there is no $+/\times$ birefringence.

To better probe $q_4$ we can define, for $p\neq0$, $y\equiv\omega^2/p^2$ yielding
\begin{equation}
 q(y)\equiv\frac{q_4}{p^4}
 =\frac{\mathcal W_s^2}{2}(y-1)^2
 +2(\mathcal W_Ey-\mathcal W_B)^2 .
 \label{eq:qy}
\end{equation}
On the positive-spectrum branch the response eigenvalues are finite, real, and nonzero. Both terms in Eq.~\eqref{eq:qy} are therefore nonnegative for real $y$, and a nonzero real characteristic exists iff both squares vanish:
\begin{align}
 q_4(k)=0,\ k_\mu\in\mathbb R,\ k_\mu\neq0
 \ \Longleftrightarrow\
 \omega^2=p^2,
 \mathcal W_E=\mathcal W_B .
 \label{eq:matching}
\end{align}
For $p=0$,
$q_4(\omega,0)=(\mathcal W_s^2/2+2\mathcal W_E^2)\omega^4$,
so no additional nonzero real characteristic occurs.

Combining Eq.~\eqref{eq:matching} with Eq.~\eqref{eq:DeltaH} gives the central result. For each of the three sectors above---vacuum, the symmetry-inheriting foundational multiplet, and the perfect-fluid constitutive embedding---and for $\beta\neq0$,
\begin{equation}
 q_4=0\ \text{for real }k\neq0
 \ \Longleftrightarrow\ \Delta_H =0 \Longleftrightarrow
 \dot H=\frac{K}{a^2}.
 \label{eq:Einstein}
\end{equation}
Thus the algebraic perfect-fluid constitutive operator has the same formal strict-principal cone obstruction as the vacuum and symmetry-inheriting foundational sectors: it changes the Ricci/one-form weight $\mathcal W_s$, but cannot alter the geometric mismatch between $\mathcal W_E$ and $\mathcal W_B$ because $M^{\rm pf}_{(2)}=0$.

The mismatch also has an invariant measure. Define the traceless Ricci tensor
\begin{equation}
 \RicTF_{\mu\nu}\equiv R_{\mu\nu}-\frac14 Rg_{\mu\nu}.
 \label{eq:RicTFdef}
\end{equation}
For FLRW,
\begin{equation}
 \RicTF^\mu{}_\nu\RicTF^\nu{}_\mu=3\Delta_H^2 .
 \label{eq:RicTFidentity}
\end{equation}
Completing the square in Eq.~\eqref{eq:qy} gives
\begin{align}
 y_*&=\frac{\mathcal W_s^2+4\mathcal W_E\mathcal W_B}
 {\mathcal W_s^2+4\mathcal W_E^2},\nonumber\\
 q_{\min}^{\mathbb R}
 &=\frac{2\mathcal W_s^2(\mathcal W_E-\mathcal W_B)^2}
 {\mathcal W_s^2+4\mathcal W_E^2}\nonumber\\
 &=\frac{8\beta^2\mathcal W_s^2\mathcal W_E^2\mathcal W_B^2}
 {3(\mathcal W_s^2+4\mathcal W_E^2)}
 \RicTF^\mu{}_\nu\RicTF^\nu{}_\mu .
 \label{eq:qmin}
\end{align}
Hence the strict-principal gap vanishes precisely on the pointwise Einstein locus
$\RicTF_{\mu\nu}=0$. If Eq.~\eqref{eq:Einstein} holds on an interval, FLRW is maximally symmetric. At matching,
$q_4=(\mathcal W_s^2/2+2\mathcal W_E^2)(\omega^2-p^2)^2$,
so the surviving characteristic is the repeated metric light cone. A repeated real cone does not by itself establish strong hyperbolicity~\cite{Reall:2021voz,Ali:2025ybt,Thaalba:2026eyv}.

Away from matching, the discriminant of Eq.~\eqref{eq:qy} as a quadratic in $y$ is
\begin{equation}
 \Delta_y=-4\mathcal W_s^2(\mathcal W_E-\mathcal W_B)^2<0,
 \label{eq:disc}
\end{equation}
and the exact squared-frequency roots are
\begin{equation}
 y_\pm=
 \frac{\mathcal W_s^2+4\mathcal W_E\mathcal W_B}
 {\mathcal W_s^2+4\mathcal W_E^2}
 \pm
 \frac{2\ii\mathcal W_s(\mathcal W_B-\mathcal W_E)}
 {\mathcal W_s^2+4\mathcal W_E^2}.
 \label{eq:roots}
\end{equation}
For any real $p\neq0$, $\omega=\pm|p|\sqrt{y_\pm}$ is therefore complex whenever $\Delta_H\neq0$ in all three sectors covered by Eq.~\eqref{eq:Einstein}. With the convention $e^{-\ii\omega t}$, the four temporal roots contain paired growing and decaying branches. This is a strict-principal hyperbolicity obstruction in the closed TT sector, not merely a shifted real gravitational-wave speed.

\emph{Vacuum inflationary consequence and nonlinear trace-log origin.---}
The cone theorem above is not restricted to vacuum; the present specialization is. For the know flat vacuum branches with $H\neq0$~\cite{Thattarampilly:2025krv}, Eq.~\eqref{eq:Einstein} becomes
\begin{equation}
 \epsilon\equiv-\frac{\dot H}{H^2}=0.
 \label{eq:epsilon}
\end{equation}
Thus every evolving vacuum slow-roll or phantom-like branch with $\epsilon\neq0$ fails the strict-principal real-characteristic condition derived here. It does not invalidate the background dynamics of~\cite{Thattarampilly:2025krv}, but it means that standard GR tensor relations cannot be justified without an independent perturbation analysis, which that work explicitly left open.

The departure from the GR cone is quantitative. At $y=1$ one has
\begin{equation}
 q(1)=2(\mathcal W_E-\mathcal W_B)^2
 =8\beta^2\mathcal W_E^2\mathcal W_B^2\epsilon^2H^4
 \quad(K=0),
 \label{eq:GRresidual}
\end{equation}
so the metric-cone residual is nonzero for every $\epsilon\neq0$. For vacuum flat FLRW define $x=2\beta H^2$. Keeping the curvature dependence exact but expanding in slow evolution gives
\begin{align}
 \operatorname{Im}y_\pm
 &=\pm\frac{x(2-3x)}{10x^2-14x+5}\,\epsilon
 +\Order(\epsilon^2),\label{eq:Imy}\\
 \frac{|\operatorname{Im}\omega|}{|p|}
 &=\frac{x(2-3x)}{2(10x^2-14x+5)}|\epsilon|
 +\Order(\epsilon^2),
 \label{eq:Imomega}
\end{align}
where the second expression follows by expanding the square root about the repeated null root. Slow evolution suppresses the complex splitting continuously but never restores a real characteristic unless $\epsilon=0$.

This sensitivity is absent in the background-independent curvature-squared Hessian around Minkowski space. Expanding
\begin{equation}
 -\Tr\ln(\tI-\beta\tR)
 =\sum_{n\ge1}\frac{\beta^n}{n}\Tr(\tR^n)
\end{equation}
shows why: the quadratic-curvature Hessian contains $(\delta\tR)^2$ with no background curvature, whereas the cubic term first contains $\bar\tR(\delta\tR)^2$ and therefore first distinguishes the FLRW $E/B$ response coefficients. This splitting alone does not yet remove the metric-cone characteristic: the cubic-truncated TT polynomial retains an exact factor $(y-1)$. Writing $\theta_A=1-\beta r_A$ for the curvature eigenvalues appearing in Eq.~\eqref{eq:geometricTheta}, the first lifting of that root in the curvature expansion occurs at quartic order, since
\begin{align}
 q(1)&=2\beta^2(r_E-r_B)^2+\Order(\beta^3), \\
 \Delta_y&=-4\beta^2(r_E-r_B)^2+\Order(\beta^3),
 \label{eq:curvatureOrder}
\end{align}
for the normalized polynomial~\eqref{eq:qy}; restoring the overall $\beta^2$ suppressed in Eq.~\eqref{eq:q4}, the metric-cone residual therefore starts at order $\beta^4$. The exact trace-log resums this nonquadratic response to all orders, with no nonzero $|\Delta_H|$ threshold at strict-principal order. It is also distinct from the finite-mass extra spin-two branch found around Minkowski space~\cite{Pereira:2026mjk}; curing or reinterpreting that branch would not remove Eq.~\eqref{eq:matching} at fixed $\beta\neq0$.

\emph{Microscopic matter: a one-multiplet rank obstruction.---}
We now ask whether the foundational multiplet $\ket{\Phi}$ can provide the needed isotropic response. Two logically independent requirements are useful.

First, full principal isotropy means that the family of frozen Douglis--Nirenberg (DN) principal symbols~\cite{Douglis1955InteriorEF} of the coupled metric--matter equations is $SO(3)$-covariant, $P(Rk)=D(R)P(k)D(R)^{-1}$. After choosing $k^\mu=(\omega,0,0,p)$, the residual symmetry is the $SO(2)$ little group about the propagation axis. The TT variables $h_\pm=h_+\pm\ii h_\times$ carry helicity $\pm2$, while scalar, one-form, two-form, and non-TT metric perturbations carry only helicities $0,\pm1$. Little-group equivariance therefore forbids principal mixing of the TT block with those sectors. This establishes the closure needed for Eq.~\eqref{eq:q4} without inferring the vanishing of any particular background spurion from a schematic variation.

Second, composite-response isotropy requires the background mixed matter endomorphisms
$\mathbb M_{(p)}\equiv\bar M_{(p)}\bar g_{(p)}^{-1}$ to be rotationally invariant in the one- and two-form sectors, where $\bar g_{(p)}$ is the corresponding background form-metric block. For one multiplet
\begin{equation}
 M_{(p)}=\ket{v_p}\bra{v_p}
 +(m^2+\xi R)\ket{\Phi_p}\bra{\Phi_p},
 \ v_p=(\Dcal\Phi)_{(p)},
 \label{eq:rankform}
\end{equation}
Here $\Phi_p$ is the $p$-form component of $\Phi$. Thus $\operatorname{rank}M_{(p)}\le2$ independently of signs or complex amplitudes.

On $\Lambda^1\simeq\mathbf1_t\oplus\mathbf3_s$, rotational invariance gives
\begin{equation}
 \mathbb M_{(1)}^{\rm iso}=a\oplus bI_3.
\end{equation}
If $b\neq0$ the spatial block has rank three, incompatible with the one-multiplet rank bound. Hence $b=0$: one multiplet cannot furnish a nonzero isotropic spatial one-form composite response. On
$\Lambda^2\simeq\mathbf3_E\oplus\mathbf3_B$, Schur's lemma gives
\begin{equation}
 \mathbb M_{(2)}^{\rm iso}=A_{2\times2}\otimes I_3,
 \quad
 \operatorname{rank}\mathbb M_{(2)}^{\rm iso}=3\operatorname{rank}A
 \in\{0,3,6\}.
 \label{eq:ranktheorem}
\end{equation}
Again rank $\le2$ forces
\begin{equation}
 {\mathbb M_{(2)}=0.}
 \label{eq:M2zero}
\end{equation}
Therefore a one-multiplet background satisfying both full principal isotropy and composite-response isotropy has no matter correction to the isotropic spatial one-form or bivector responses relevant to Eq.~\eqref{eq:q4}; it falls within the same cone-matching theorem as the symmetry-inheriting foundational sector.

The counting also identifies the first possible escape. For $N$ additive foundational-type multiplets,
\begin{equation}
 \operatorname{rank}M_{(p)}^{(N)}\le2N.
\end{equation}
Thus $N=2$ is the first multiplicity not excluded by rank counting for a nonzero isotropic rank-three spatial one-form or bivector response, while a full rank-six bivector response requires $N\ge3$. These are necessary counting conditions only: the individual multiplets must also assemble into an $SO(3)$-equivariant principal coefficient set, and new helicity-two matter combinations can enlarge the tensor determinant. Multi-form cosmologies provide precedent for isotropic collective structures~\cite{Koivisto:2009sd,Ajith:2022wia,Horii:2025jen}, but the characteristic problem must then be recomputed.

The amount of response required is fixed. For an $E/B$-diagonal isotropic extension, let $\bar m_E$ and $\bar m_B$ denote the eigenvalues of the mixed bivector matter endomorphism $\mathbb M_{(2)}$ on $\mathbf3_E$ and $\mathbf3_B$. The corresponding eigenvalues of $\bar\tTheta$ are
\begin{align}
 \theta_E&=1+\alpha\bar m_E-2\beta(H^2+\dot H),\nonumber\\
 \theta_B&=1+\alpha\bar m_B-2\beta(H^2+K/a^2).
 \label{eq:matterThetaEB}
\end{align}
with inverse responses $\mathcal W_E=\theta_E^{-1}$ and $\mathcal W_B=\theta_B^{-1}$. Since cone matching is $\mathcal W_E=\mathcal W_B$, equivalently $\theta_E=\theta_B$ on the positive-spectrum branch, a non-Einstein background requires
\begin{equation}
 {
 \bar m_E-\bar m_B
 =\frac{2\beta}{\alpha}\left(\dot H-\frac{K}{a^2}\right) }
 \quad(\alpha\neq0).
 \label{eq:compensation}
\end{equation}
Equation~\eqref{eq:compensation} is a microscopic design condition, not a sufficiency theorem: full principal isotropy, positivity of the trace-log branch, and hyperbolicity of the enlarged tensor system remain to be checked.

\emph{Auxiliary G-field reformulation: constrained equivalence.---}
All results above were obtained in the foundational metric--matter theory. We now show that the auxiliary G-field representation of Ref.~\cite{Bianconi:2024aju} gives exactly the same strict-principal TT dynamics on its constrained branch. Introduce independent auxiliaries $\tTheta_{\rm a}$ and $\Gfield$ only for this purpose,
\begin{equation}
 \widetilde{\mathcal L}
 =-\Tr\ln\tTheta_{\rm a}
 -\Tr\!\left[\Gfield(\tG\tg^{-1}-\tTheta_{\rm a})\right].
 \label{eq:canonical}
\end{equation}
Variation with respect to $\Gfield$ and $\tTheta_{\rm a}$ is purely algebraic and gives $\tTheta_{\rm a}=\tG\tg^{-1}=\tTheta$, $\Gfield=\tTheta_{\rm a}^{-1}=\Wresp$ and substituting back into Eq.~\eqref{eq:canonical} removes the multiplier term identically and returns $-\Tr\ln(\tG\tg^{-1})$, i.e. the foundational Lagrangian~\eqref{eq:originalAction}. The equivalence is therefore exact at the action level for the algebraically constrained auxiliary system considered here.

The same statement can be checked at the principal-symbol level without assuming the answer. Linearize Eq.~\eqref{eq:canonical}, eliminate $\delta\tTheta_{\rm a}$ algebraically, and denote
$\mathfrak g\equiv\delta\Gfield$ and
$\mathsf X\equiv\delta(\tG\tg^{-1})$. Let
$\mathcal C^{(2)}(k)$ be the two-derivative linear curvature map defined by
$\mathcal C^{(2)}(k)h\equiv\delta(\tR\tg^{-1})_{\rm TT}^{(2)}$,
and let $P_{hh}^{(2)}(k)$ denote the remaining two-derivative TT metric block. Here $h=(h_+,h_\times)$ and $\dagger$ denotes the adjoint with respect to the quadratic trace pairing.
In the closed TT sector one obtains the mixed frozen system
\begin{equation}
 \begin{pmatrix}
 P_{hh}^{(2)}&\beta\mathcal C^{(2)\dagger}\\
 \beta\mathcal C^{(2)}&-\mathcal A
 \end{pmatrix}
 \binom{h}{\mathfrak g}=0,
 \quad
 \mathcal A[X]=\bar\Gfield^{-1}X\bar\Gfield^{-1}.
 \label{eq:auxblockmain}
\end{equation}
On the positive-spectrum branch $\mathcal A$ is invertible. Exact elimination of $\mathfrak g$ gives
\begin{equation}
 P_{hh}^{(2)}+
 \beta^2\mathcal C^{(2)\dagger}\mathcal A^{-1}\mathcal C^{(2)},
 \quad
 \mathcal A^{-1}[X]=\bar\Wresp X\bar\Wresp.
 \label{eq:auxschurmain}
\end{equation}
The degree-four term in Eq.~\eqref{eq:auxschurmain} is exactly the trace-log Hessian~\eqref{eq:Hessian}. The apparent difference in differential order is therefore only representational: the foundational metric equation contains the fourth-order reduced operator, whereas the auxiliary formulation keeps a mixed lower-order system whose algebraic elimination regenerates it. Moreover, the DN principal determinant factorizes by the corresponding Schur complement, so the constrained auxiliary and foundational formulations have the same TT characteristic set whenever the algebraic block is invertible.

\emph{Interpretation and regime of validity.---}
An important point is to assess whether the obtained results are scale dependent or admit an EFT interpretation. The homogeneous scaling $q_4(\lambda\omega,\lambda p)=\lambda^4q_4(\omega,p)$ means that there is no momentum threshold intrinsic to the strict-principal polynomial: ``ultraviolet'' refers to its asymptotic role in the unreduced mixed-order equations. At $\beta=0$ the quartic \emph{TT} contribution itself vanishes and the TT differential order drops, so lower-derivative terms determine propagation. For fixed $\beta\neq0$ the degree-four term dominates the formal $|p|\rightarrow\infty$ limit. In weak-curvature/Minkowski scaling the two- and four-derivative pieces behave parametrically as $\beta p^2$ and $\beta^2p^4$, placing their crossover near $|p|\sim|\beta|^{-1/2}$ up to order-one response factors. If the trace-log theory is only an EFT with a cutoff below that scale, the fourth-order regime need not be physically sampled; the obstruction should then be interpreted as a property of the unreduced continuum completion rather than automatically as a finite-frequency observational prediction.

\emph{Conclusion and discussion.---}
We found that the exact degree-four TT symbol on FLRW is the sum of a Ricci-response square and a Riemann-response square. In vacuum, for the symmetry-inheriting foundational single multiplet, and for the algebraic perfect-fluid constitutive embedding, the two squares have a common nonzero real zero only when \(\dot H=K/a^2\). Equivalently, the traceless Ricci tensor \(\RicTF_{\mu\nu}\) vanishes and the repeated metric light cone is recovered. Away from this locus the real-axis symbol develops the positive gap~\eqref{eq:qmin}, the squared characteristic roots~\eqref{eq:roots} are complex, and the closed TT principal system is nonhyperbolic. For this perfect-fluid constitutive embedding, the result is therefore not merely analogous to the vacuum case: because the operator has no bivector block, it inherits the same \(E/B\) response mismatch directly. Consequently, under the standard non-EFT interpretation of GfE, generic non-Einstein FLRW solutions in these settings are ruled out as viable cosmological backgrounds, since their gravitational-wave sector fails to support physically propagating high-frequency tensor modes.

The vacuum slow-roll specialization gives a near-de Sitter complex splitting linear in \(\epsilon\), while the microscopic rank theorem identifies what must change if GfE is to accommodate generic non-Einstein FLRW evolution within an isotropic matter sector. A single foundational multiplet cannot furnish a nonzero rotationally invariant spatial one-form or bivector composite response under the stated assumptions; \(N=2\) is only the first multiplicity not excluded by rank counting, and Eq.~\eqref{eq:compensation} fixes the required background \(E/B\) response difference. These results therefore indicate that viable tensor propagation on generic cosmological backgrounds may require a richer matter sector, or a more general collective realization of the foundational GfE degrees of freedom, capable of compensating the curvature-response splitting.

More broadly, this analysis illustrates how perturbative consistency can place nontrivial constraints on the microscopic structure of approaches in which gravitational dynamics are tied to thermodynamic or information-theoretic principles. Reproducing Einstein-like background evolution does not by itself guarantee a viable Lorentzian perturbation theory: in GfE, the tensor characteristic structure probes response information that is invisible to the homogeneous background equations. Future work should determine whether multi-multiplet or other collective realizations of the GfE matter sector can generate the compensating bivector response required by Eq.~\eqref{eq:compensation} while preserving positivity and a well-posed coupled dynamics, and should extend the present strict-principal analysis to the complete finite-frequency metric--matter perturbation system.

\begin{acknowledgments}
We thank Jess Rutschi for useful discussions. Supported by FCT through the research grant UID/04434/2025.
\end{acknowledgments}

\paragraph*{Data availability.---}
No data were created or analyzed in this study.

\bibliographystyle{apsrev4-2}
\bibliography{biblio_GfE}

\appendix
\section{Technical checks}\label{app:checks}

\emph{Direct TT contractions.---}
Here $\delta R_{(1)}$ denotes the mixed Ricci perturbation acting on $\Lambda^1$ and
$\delta R_{(2)}$ the mixed Riemann operator acting on $\Lambda^2$.
At principal TT order $\delta R=\delta R_{00}=\delta R_{0i}=0$ and
$\delta R^i{}_j=-\tfrac12(\omega^2-p^2)h^i{}_j$. With an isotropic spatial one-form response,
\begin{equation}
 \Tr_1(\bar\Wresp_{(1)}\delta R_{(1)}
 \bar\Wresp_{(1)}\delta R_{(1)})
 =\frac{\mathcal W_s^2}{4}(\omega^2-p^2)^2h_{ij}h^{ij}.
 \label{eq:oneformApp}
\end{equation}
For a $+$ wave propagating along $z$, use the flattened independent-pair basis
$(01,02,03,23,31,12)$. With the foundational convention
\[
 [g_{(2)}]_{\mu\nu\rho\sigma}
 =\tfrac12(g_{\mu\rho}g_{\nu\sigma}-g_{\mu\sigma}g_{\nu\rho}),
\]
the independent-pair flattening carries the corresponding $2!$ factor~\cite{Bianconi:2024aju}. The resulting foundational two-form trace is
\begin{equation}
 \Tr_2(\bar\Wresp_{(2)}\delta R_{(2)}
 \bar\Wresp_{(2)}\delta R_{(2)})
 =(\mathcal W_E\omega^2-\mathcal W_Bp^2)^2h_{ij}h^{ij}.
 \label{eq:bivectorApp}
\end{equation}
Equations~\eqref{eq:oneformApp} and \eqref{eq:bivectorApp} give Eq.~\eqref{eq:q4} for $e^A_{ij}e^{B\,ij}=2\delta^{AB}$.

For a closed TT block, matter introduces no additional two-derivative metric insertion. The metric variation of the Hodge--Dirac operator obeys $\delta_g\Dcal=\Order(\partial h)$; variations of $\tg^{-1}$ are algebraic; and the only two-derivative metric term generated by $(m^2+\xi R)|\Phi\rangle\langle\Phi|$ is
\begin{equation}
 \left.\delta_gM_{(p)}\right|_{2\partial}
 =\xi\,\delta R\ket{\bar\Phi_p}\bra{\bar\Phi_p},
\end{equation}
which vanishes in strict-principal TT because $\delta R=0$.

\end{document}